# Computer-assisted access to the kidney

P. MOZER[1,2], A. LEROY[2], Y. PAYAN[2], J. TROCCAZ[2],
E. CHARTIER-KASTLER[1], F. RICHARD[1]

[1] Department of Urology, Pitié-Salpetriere Hospital,
Pierre et Marie Curie University (Paris VI) Paris, France.

[2] TIMC Laboratory, IN3S, Faculté de Médecine
Domaine de la Merci, 38706 La Tronche cedex, France.

**3800 words**

# Abstract :

Objectives: The aim of this paper is to introduce the principles of computer-assisted access to the kidney. The system provides the surgeon with a pre-operative 3D planning on computed tomography (CT) images. After a rigid registration with space-localized ultrasound (US) data, preoperative planning can be transferred to the intra-operative conditions and an intuitive man-machine interface allows the user to perform a puncture.

Material and methods: Both CT and US images of informed normal volunteer were obtained to perform calculation on the accuracy of registration and punctures were carried out on a kidney phantom to measure the precision of the whole of the system.

Results: We carried out millimetric registrations on real data and guidance experiments on a kidney phantom showed encouraging results of 4.7mm between planned and reached targets. We noticed that the most significant error was related to the needle deflection during the puncture.

Conclusion: Preliminary results are encouraging. Further work will be undertaken to improve efficiency and accuracy, and to take breathing into account.

# Introduction

Percutaneous access to the kidney is a challenging technique that meets with the difficulty to reach rapidly and accurately a target inside the kidney. For example, in case of percutaneous nephrolithotomy (an intervention performed to remove stones from the kidney), it is shown that optimizing the progress of the puncture, by targeting a fornix [1], allows to decrease the risk to perforate large vessels inside this organ.

Nowadays, in clinical practice puncture guidance is performed under fluoroscopic and/or echographic imaging, each of which presents drawbacks. Fluoroscopy provides limited two-dimensional (2D) information on localization and involves patient and operator irradiation, whereas echography mostly gives fuzzy images of both target and puncture trajectory.

To minimize the drawbacks of these modalities, some teams investigated the use of computed tomography (CT) [2] or magnetic resonance imaging [3], but these tools are time consuming, are not ergonomic, and do not take the movements of the kidney into account.

To our knowledge, only one device has entered the clinical field to help the physician to perform kidney puncture. This system called PAKY [4] is based on visual servoing. The operator directly localizes a target on fluoroscopic images, and then a robot performs the puncture under human control. This action is executed during the patient apnoea. Work is in progress to automate the puncture from CT [5] or fluoroscopic images [6], but moving the C-arm in two different positions to locate the target is still necessary, and only the collecting system can be reached.

We introduce the principles of computer-assisted access to improve the current clinical practice. This system provides the surgeon with an accurate pre-operative three-dimensional (3D) planning on CT images and, after a registration with space-localized echographic data, would help him to perform the puncture through an intuitive user interface.

# Materials and methods

## *The approach*

The general approach consists in the following steps:

- A 3D pre-operative model is reconstructed from abdominal CT images (global shape of the kidney, collecting system, ribs, spine, lungs and skin).
- The surgeon uses this model to define a planning by selecting two points, a target and an entry point, which define the needle trajectory.
- Just before puncture, intra-operative ultrasonic (US) images are collected to get a set of 3D points located onto the kidney surface. As echography is used here like a tool to locate the surface of the kidney, it is not necessary for the target to be visible in the ultrasound images.
- This set of 3D points is matched onto the preoperative model of the kidney, by the mean of a 3D/3D rigid registration technique. The matching transformation applied to the planned trajectory allows transferring it to the operating room (OR) conditions and guarantees its correct execution. The position of the surgical tool is known in real-time during the surgical action thanks to a localizer and compared to the planned trajectory. Therefore, no further image acquisition is needed for this guiding phase.

The three main stages, namely planning, registration and guidance, are described below in details.

## Planning phase

Pre-operative 3D data are collected. In these data, two kinds of relevant anatomical structures have to be segmented. Structures that will participate in the planning: the target (often in the pyelocalyceal system) and structures such as lungs or ribs are called "planning structures".

Structures which are used for registration are called "reference structures": e.g. the kidney surface. 3D representations of these two kinds of structures are obtained in the pre-operative coordinate system.

By taking the information provided by these structures into account, the planning phase allows the selection of a needle trajectory and a target position.

## Registration phase

US data of the reference structures are collected just before the puncture. The echographic probe is equipped with localisation features (Fig 1) which are tracked in real-time using a localizer. Each time an image is recorded, the 6 position parameters of the probe are also recorded thus localizing the 2D ultrasound image in the 3D space. We call this device "2.5D echography". Thanks to the image position, the segmented structures are also localized in the 3D space, thus allowing building a 3D representation of the reference structures in the intra-operative coordinate system.

Let us mention that this resulting representation may be a sparse set of data. Those data, indeed, are used during registration to compute the geometric transformation between the set of 3D points and the pre-operative model of the kidney. Therefore, a complete and homogeneous echographic reconstruction of the whole kidney surface is not necessary. At that point, planning data can be mapped to the intra-operative conditions using this transform.

**Guidance phase**

We chose a passive system based on surgical instrument tracking capabilities providing information to compare the executed trajectory to the planned trajectory (Fig 2) but other kinds of guiding systems [7] might be used to reach the target position through a planned trajectory.

## Experiments

Most of the development stage relied on both CT and US images of a healthy subject. In parallel, in order to check our numerical results we also made 3 punctures on a right kidney of a phantom. The main purpose of this preliminary work was to evaluate the feasibility of image-based guidance, provided that the image modalities were determined by conventional procedures. We aimed at being able to quantitatively evaluate the algorithms that register CT with "2.5D echography" in rather realistic conditions.

In the first stage of this study, it was assumed that the key steps of the protocol, namely image acquisitions and guidance, could be executed at the same moment in the respiration cycle. This assumption will be discussed later.

**CT data acquisition**

Pre-operative data were acquired from an informed healthy volunteer with normal urinary tract with a CT scan "Light Speed Ultra" from General Electric. After an intravenous bolus injection with 120 mL contrast medium (300 mgI/mL at 4 mL/s), all scans were taken at 120 kV and 220 mA.s.
For the first phase scan (noted SE1), a delay of 15 s was used; this scan extended from the coeliac axis superiorly to the aortic bifurcation inferiorly, and was taken with 3 mm collimation and a 5 mm/s table speed. This acquisition provided a scan time of less than 30 s at full inspiration and gives accurate information of the surface of the kidney and on the parenchyma (Fig 3). Overlapping images were reconstructed at a mean (range) of 2 (1.5  2.5) mm intervals.
The final component was a pyelocalyceal phase scan (noted SE2) at 180 s delay with a 5 mm slice thickness, 5 mm/s table speed and 2.5 mm reconstruction interval. This acquisition gives accurate information on the collecting system which is often the target to reach in clinical practice but the surface of the kidney is fuzzy (Fig 4).

**Surface registration between each CT scan acquisition**

Because the patient had breathed and sometimes moved between these two acquisitions, kidneys are not on the same position in the CT volumes. In order to have the whole information in a single representation a 3D/3D registration is carried out between each phase scan of each kidney based on their surface (we assume that there is no deformation of the kidney between each acquisition).
For each phase scan, the external surface of the kidney was segmented using derivatives methods (Nabla's 3D watershed [8] from Generic Vision[1]) and the registration was performed with the Analyze®[2] software. The registration method is a rigid surface matching algorithm using a distance map.

---

[1] Glentevej 67, DK-2400 Copenhagen NV, Denmark
[2] AnalyzeDirect, Inc. 11425 Strang Line Road. Lenexa, KS, 66215 USA. www.analyzedirect.com

"Planning structures" such as ribs, spine, lungs, skin and collecting system were segmented using together Nabla's 3D watershed and threshold. The generated model makes possible to carry out a planning through an intuitive 3D interface (Fig 5).

In a second stage – that simulates intra-operative procedures – the "2.5D echographic" acquisition is performed during an apnoea at the end of an inspiration. The echographic system was a HITACHI-EUB450 with a 3.5 MHz probe. The optical localizer was a passive Polaris system from Northern Digital Inc[3].
We acquired 200 images at 3 images per second, by a lateral echographic window, in both transversal and longitudinal orientations. We do not need so many images for registration but it was decided to acquire the most images possible during an apnoea to test the registration precision and to set up an optimal strategy for image acquisition.

**Surface registration between CT and US data**
The registration makes use of a surface matching algorithm using a distance map recorded in an octree-spline data structure [9]. This data structure is computed from the densest representation, namely the pre-operative CT model in our case. The algorithm iteratively moves the sparse representation relatively to the dense one and computes the parameters that minimize the distance function between the two representations. At a starting point, only rigid matching has been used, which explains that intrinsic deformation of the kidney are not taken into account. We assume that this deformation between CT and US data is very small and can be neglected in this first development stage. Because data were acquired in conditions where no gold standard was available – in other words, it was not possible to know the exact transform between CT and echographic data – two different tests were used to evaluate the registration.

The first test named "repeatability test" consists in running the registration algorithm from several initial relative positions of the CT and echographic representations and to observe the repeatability of the computed transform. Such a test informs on the presence of local minima in the vicinity of the solution; such minima can result in misregistrations of data. The value of the residual mean square (rms) after registration is also an indicator of the registration accuracy: a large rms would mean an inaccurate registration or mismatched data.

The second test is named "closed-loop accuracy test". The idea is to compare 3 related registrations. Let *CT1* and *CT2* be two CT meshes of the kidney, where *CT2* is a transformed *CT1* (e.g. 50mm in translation and 10° on each rotation angle). Let *US* be an echographic cloud of points of the same organ. *M12*, *M1U* and *M2U* are the mono- or multi-modal transforms betweens the exams. Our closed-loop test consists in evaluating $\|\delta M \times CT_1 - CT_1\|$ with $\delta M = M1U^{-1} \times (M2U \times M12)$ (Fig 6). The registration is perfect if $\|\delta M \times CT_1 - CT_1\| = 0$.

---

[3] 103 Randall Drive, Waterloo, Ontario, Canada

### Puncture of phantom

We used a phantom from CIRS[4] (model 057). It is a 3D abdominal phantom which mimics human tissues under ultrasound and CT. Two CT exams were made, the first one for the planning and the second one after punctures. A rigid registration between each CT was made to compare the planned targets with the reached targets. For the punctures, we used urological needles from Boston Scientific (18-gauge, 200mm long). The tip of the needle was localised in space thanks to a rigid body placed on its proximal part (Fig 7).

---

[4] 2428 Almeda Avenue, Suite 212 Norfolk, VA 23513

# Results

### *Pre-operative segmentation of CT scan*

Using Nabla's watershed algorithm, less than 5 minutes are necessary to segment the kidney surface for each CT scan. The accuracy, estimated visually by 3 different operators, was considered as being millimetric for the SE1 acquisition. For the second acquisition SE2, the segmentation was less accurate because the limit of the surface was fuzzy especially where there is a contact between the kidney and other organs (like the liver for the right kidney and the spleen for the left kidney).

The precision of the registration was estimated visually as excellent by superimposing the segmented data (Fig 8). To quantify the precision of this registration, we compared the position of kidney's centroid in SE1 and SE2. We found a distance of less than one millimetre between each centroid.

Therefore, after this phase of registration, kidney structures are available in the same CT coordinate system (Fig 9).

### *Echographic segmentation*

From the 200 recorded ultrasound images we selected 10 of them. The kidney surface was manually segmented with approximately 40 points on each image to obtain a model of 434 points (Fig 10). We noticed that a dense and homogenous cloud of points was suitable for the registration phase. However, the user may focus on some curved regions that will avoid local minima during registration.

### *Repeatability test*

A transform is represented as one translation vector ($T_x$, $T_y$ and $T_z$) and 3 rotation angles ($\Psi$, $\theta$, and $\Phi$). Tab 1 shows the obtained results for six initial positions. The deviations between the final position and the 6 initial attitudes go up to 30 mm in translation and 20° in rotation. Beyond those values, local minima are quasi-systematically found. In practice, these values can be easily reached by manual or semi-automatic initialization through anatomical landmarks. The typical value of the rms is less than one millimetre. The results are thus fairly good.

### *Closed-loop accuracy test*

The closed loop accuracy test was performed on a set of combined registration data. Our results are: $\|\delta M \times CT_1 - CT_1\| = 1.2 mm \pm 0.4$. Let us remind that it is a cumulative error of 2 consecutive registrations.

### *Puncture of phantom*

As the phantom is made of highly heterogeneous material, the pre-operative segmentation was painful. Nevertheless, the registration between CT data and ultrasound data was correct (Fig 11). Our results showed that the tip of the needles was 4.7 mm away on average from their target (tab 2). This result gives the precision of the whole of the system but also includes the registration error between the pre and post-operative CT scan.

# Discussion

## *Accuracy issues*

Many sources of errors can be mentioned to explain our results:

- CT and echographic calibration: the image parameters (scale, mm/pixel ratios and geometric relationship between the probe and the localizing features) determined by calibration procedure may introduce errors. The typical rms values after calibration are about 1 mm for echographic acquisition. Moreover, US image reconstruction performed at constant velocity may result in distorted representations of the organs. Modeling these distortions has not been integrated in this study.
- Echographic acquisition: the estimated time elapsed between the recordings of the rigid transformation and of the image is 70ms, which induces an error of 0.7mm at a 10mm/s motion.
- CT and echographic data segmentation: considering the CT pixel/mm ratio of 0.6 mm and the size and the quality of images, it looks reasonable to consider that a 1 to 2 pixels error results from segmentation. This corresponds to a 0.6 to 1.2 mm error. The same observation can be made with US images.
- Registration: the registration error is directly related to the quality of data. Tests of the registration algorithms performed on rigid phantoms demonstrated sub-millimeter accuracy. In the presented experiments, distortions of the imaging modalities (US in particular) may degrade the results.

However, among all possible sources of errors, the main inaccuracies certainly come from the symmetrical shape of the kidney's phantom which introduces potential indeterminations and from the deformation of the needle during puncture. Indeed, during the puncture, the deformation of the needle was visually very important. We think that it could be judicious to use a more rigid needle but we wish to validate our system with instruments used in clinical routine. Another solution would be to use a magnetic localizer to determine the position of the tip of this needle.

## *Clinical applicability*

Two main approaches can be envisioned for action guidance. The first one was adopted for the present work; it considers that the motion of the kidney can be cancelled thanks to apnea conditions between echographic data acquisition and needle guidance. For Davies and al. [10] kidney movement is complex during breathing but it returns to the same place after each inspiration. We could get this information by monitoring respiratory volumes by a simple respiratory gating device.

The second approach would be track the position of the kidney and could be particularly useful to carry out a percutaneous treatment by HIFU of kidney tumours. The use of a ureteral stent equipped with an electromagnetic coil could be effective in this case.

Regarding puncture, a navigational assistance can be used to guide the surgical action, but a robot would probably make the action faster and therefore more accurate.

Intra-operative image processing tools must be developed to avoid any involvement of the user in tasks other than supervision. This is why a segmentationless registration approach is under development [11]. The method consists in optimizing a rigid 6 degree of freedom transformation by evaluating at each step the similarity (correlation ratio in particular) between the set of US images and the CT volume. This approach will be integrated to a further version of this system to suppress user involvement in intra-operative image processing. Pre-operative image segmentation is less critical; meanwhile, approaches such as proposed in [12] could be introduced to obtain a fully automatic process.

## Conclusion

In this study, the bases of a computer-aided system for percutaneous access to the kidney were presented. The aim was to evaluate the feasibility and the accuracy of each step of the process. In our study, preoperative CT data were registered to intra-operative, manually segmented ultrasound data, using a 3D/3D rigid matching. Tests on registration as well as guidance experiments were satisfactory. Nevertheless, further work will be undertaken to improve efficiency and accuracy, and to take breathing into account.

# Tables :

**Tab 1 : Repeatability test results (σ = standard deviation and $\|\sigma\|=\frac{\sigma}{Mean}$ ). Distances are in mm and angles are in °.**

|   | Test 1 | Test 2 | Test 3 | Test 4 | Test 5 | Test 6 | Mean | σ | ‖ σ ‖ |
|---|---|---|---|---|---|---|---|---|---|
| $T_x$ | 270.15 | 268.77 | 272.71 | 270.50 | 272.91 | 276.14 | 271.86 | 2.63 | 0.96 |
| $T_y$ | 466.00 | 464.90 | 461.97 | 464.61 | 463.22 | 462.83 | 463.92 | 1.50 | 0.32 |
| $T_z$ | -332.32 | -335.10 | -332.93 | -333.15 | -332.09 | -327.57 | -332.19 | 2.50 | -0.75 |
| Ψ | -85.20 | -86.33 | -82.96 | -85.59 | -83.39 | -82.13 | -84.27 | 1.67 | -1.98 |
| θ | -44.49 | -44.96 | -41.78 | -44.28 | -42.24 | -40.97 | -43.12 | 1.66 | -3.85 |
| Φ | -179.73 | -179.83 | -178.72 | -179.86 | -179.18 | -179.76 | -179.51 | 0.46 | -0.25 |
| rms | 0.69 | 0.87 | 0.70 | 0.92 | 0.77 | 0.77 | 0.79 | 0.08 |  |

**Tab 2 : Distances (in mm) between pre-operative and post-operative target (d|Pre_Op-Post-Op) for 3 punctures on right kidney.**

|   | Puncture n° 1 | Puncture n°2 | Puncture n°3 | Average |
|---|---|---|---|---|
| d|Pre_Op-Post_Op| | 6.1 | 3.3 | 4.7 | 4.7 |

# Figures:

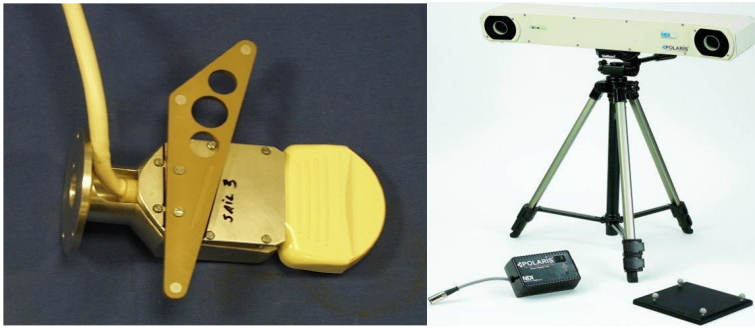

**Figure 1: Left : echographic probe with rigid body. Right : Polaris system.**

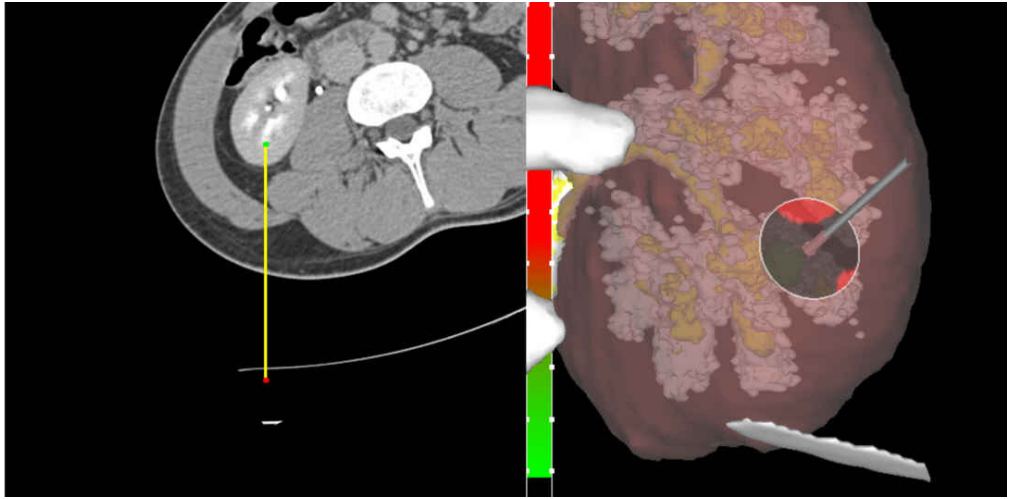

**Figure 2 : Guidance phase with navigation interface. On the left: CT reconstruction in the puncture plane. On the right: 3D interface.**

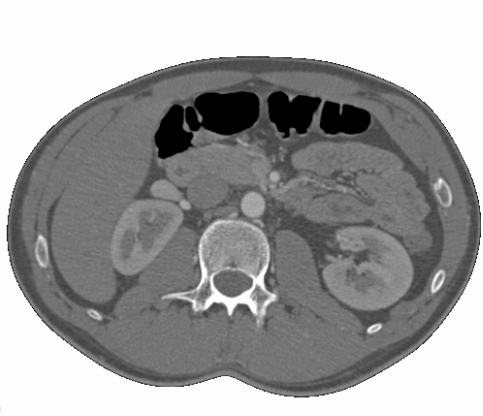

**Figure 3 : First CT scan acquisition (SE1)**

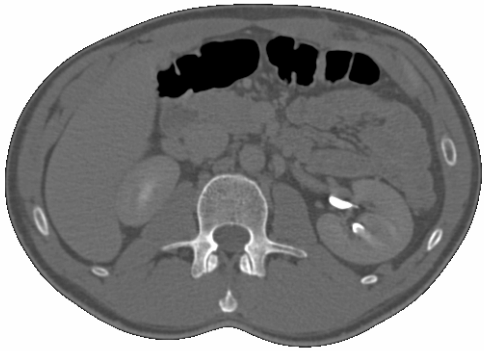

**Figure 4 : Second CT scan acquisition (SE2)**

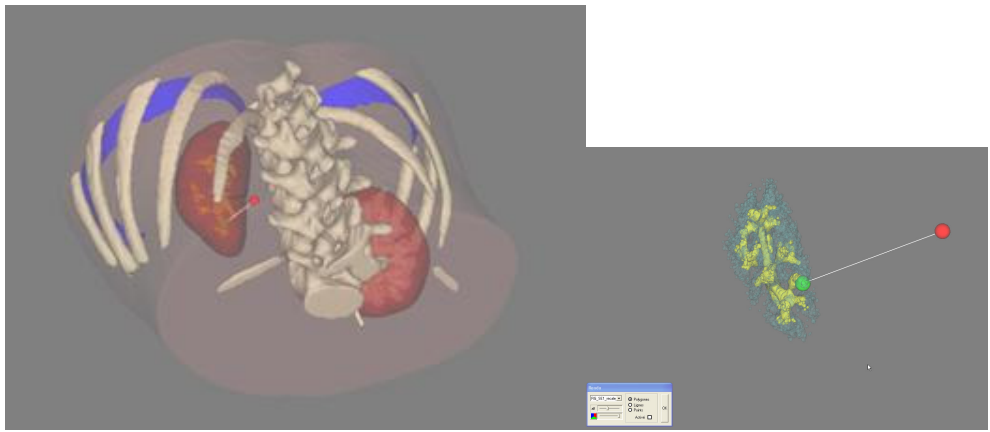

**Figure 5 : Planning 3D interface (Green: target - Red: entry point).**

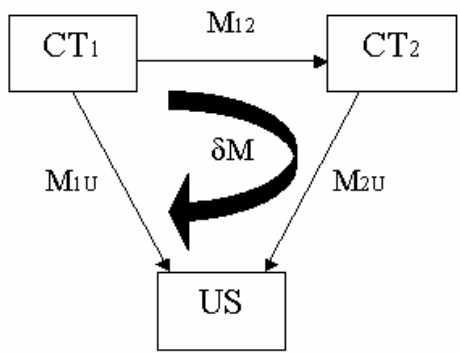

**Figure 6 : Closed-loop test.**

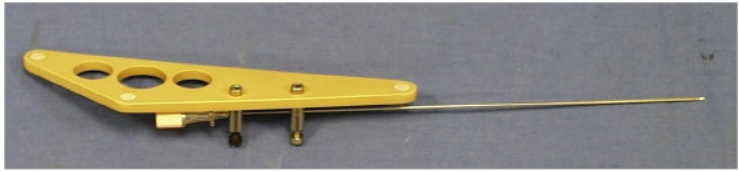

**Figure 7 : Needle equipped with a passive rigid body for position tracking.**

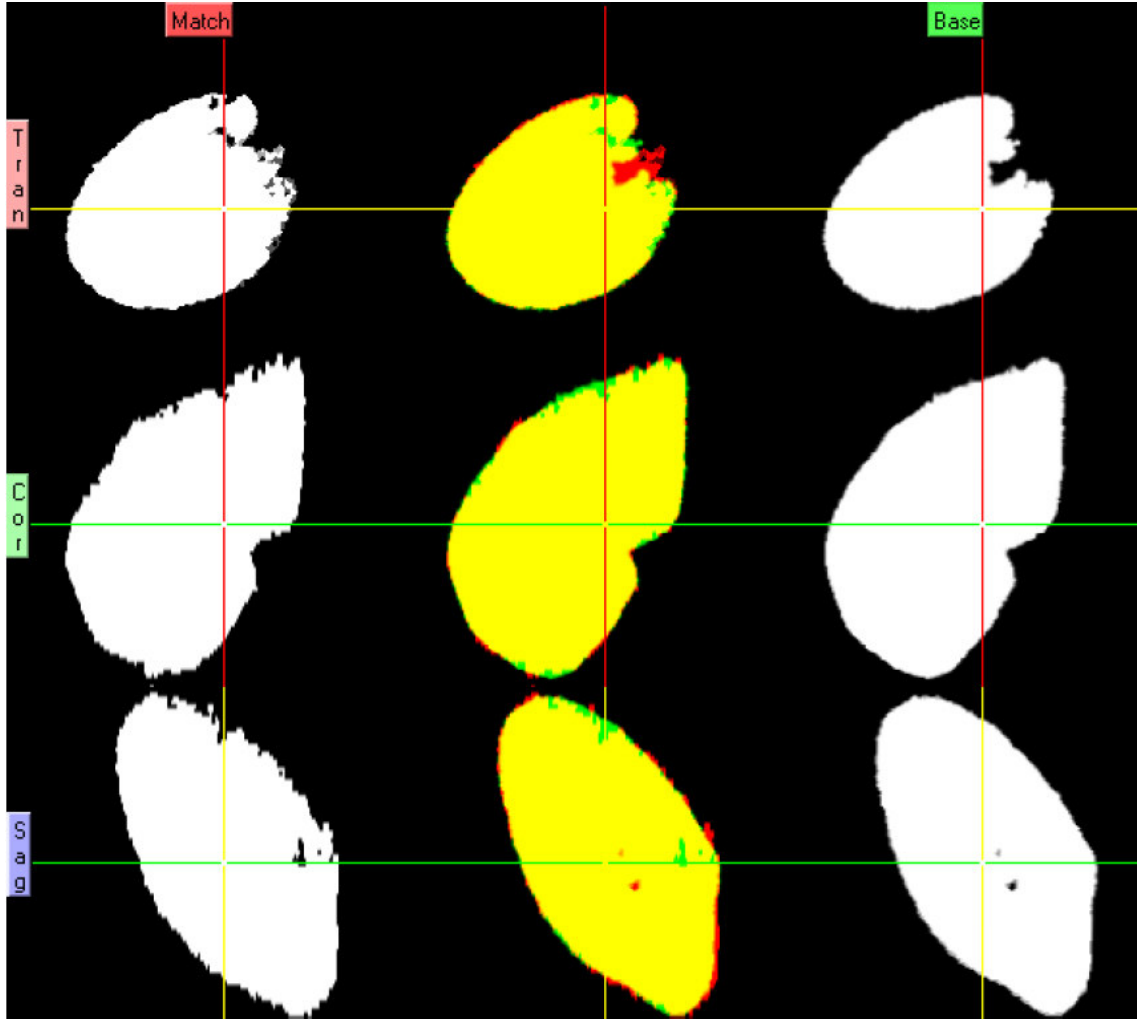

**Figure 8 : Accuracy of the registration between SE1 and SE2. Left : segmentation of SE2. Right : segmentation of SE1. Middle : yellow = SE1 U SE2. Green = SE1. Red = SE2.**

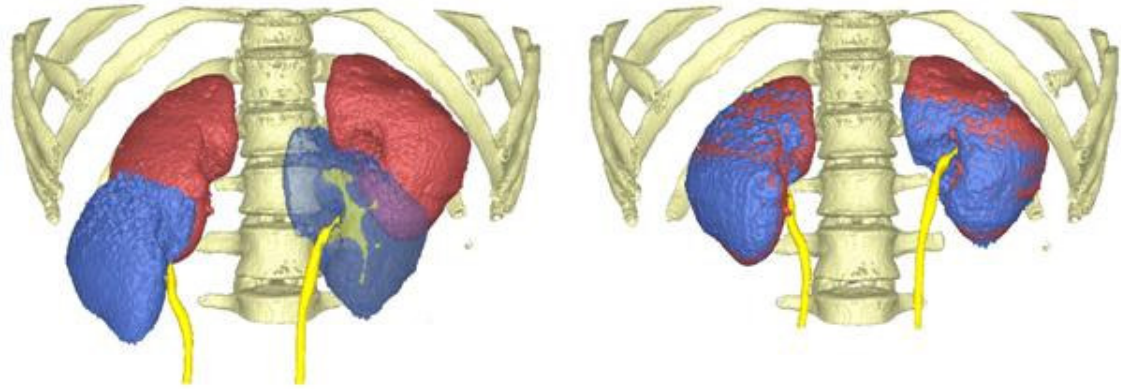

**Figure 9 : CT scan registration (Left : before registration, Right : after registration ; Red : kidney surface segmented in SE1, Blue : kidney surface segmented in SE2, Yellow : pyelocalyceal system).**

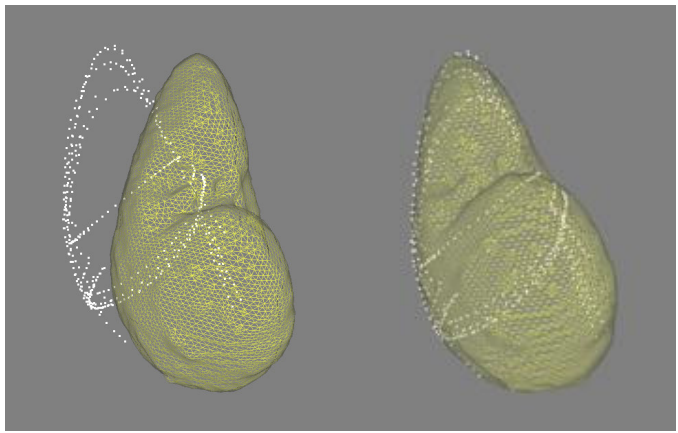

**Figure 10 : Registration of CT (yellow triangles) and US data (white points) acquired on a healthy subject. Left: before registration - right: after registration.**

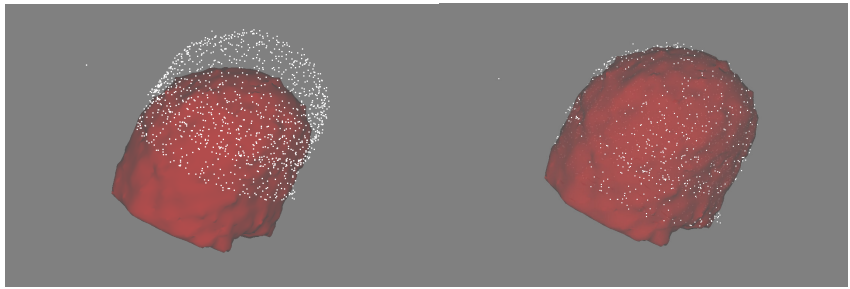

**Figure 11 : Registration of CT (red) and US data (white points) acquired on the phantom. Left: before registration - right: after registration.**

17